\documentstyle[aps,preprint]{revtex}
\tightenlines
\begin{document}

\title{Kinetic
 description of diffusion-limited reactions in random catalytic
media}

\author{G.Oshanin$^{1,}\footnote{Present address: 
Laboratoire de Physique Th\'eorique des Liquides (URA 765), 
Universit\'e Paris VI, T.16, 4 place Jussieu, 75252 Paris Cedex 05, France
}$ 
  and A.Blumen$^2$}

\address{$^1$ Centre de Recherche en Mod\'elisation Mol\'eculaire, 
Service de
 Physique Statistique, Universit\'e de Mons-Hainaut, 20 Place
du Parc, 7000 Mons, Belgium\\
$^2$ Theoretical Polymer Physics, University of Freiburg, Rheinstrasse 12, 
79104 Freiburg, Germany}

\address{\rm (Received: )}
\address{\mbox{ }}
\address{\parbox{14cm}{\rm \mbox{ }\mbox{ }
We study the kinetics of bimolecular,
catalytically-activated
reactions (CARs)
 in $d$-dimensions.
The elementary reaction act between reactants
takes place
only when these
meet
 in the vicinity of a catalytic site; such sites
are assumed to be immobile and randomly distributed in space.
For CARs we develop a kinetic formalism, 
based on 
 Collins-Kimball-type ideas; 
within this formalism
we obtain explicit expressions for the
 effective reaction rates 
and for the decay of the
reactants' concentrations.
}}
\address{\mbox{ }}

\vspace{4cm}
\address{\parbox{14cm}{\rm PACS No:  82.20.Mj; 05.40.+j; 68.10.Jy}}
\maketitle

\makeatletter
\global\@specialpagefalse

\makeatother

\section{Introduction.}

Many industrial and technological processes depend on 
catalytically-activated reactions (CAR), whose understanding, however, used to be mainly
phenomenological$^{1}$. More microscopically inclined approaches emerged only during the last decades.
Thus  much progress was made in determining how reactions are promoted by 
 specific catalytic
 substrates$^{2}$. From the point of view of many-body effects much understanding 
 was
 gained from an extensive study of the CO-oxidation on
metal surfaces$^3$.
The first
type of research$^2$ is at the molecular level and, ideally, 
leads to   $K_{el}$,  the reaction rate for
two molecules which 
meet at a catalytic site. However, 
 the results of Refs.3 demonstrate amply 
that the mere knowledge 
of $K_{el}$
is not sufficient for detemining 
 the time evolution 
of  the global reaction process. Thus the decay forms for the reactants' concentrations in CARs
often deviate strongly from the results of formal-kinetics schemes$^3$. We note that such deviations
are not encountered in CARs only, but are widespread$^{4-6}$, being due to many-body effects,
 often associated with fluctuations in the spatial distributions of the reacting species.
  Consequently, a robust, reliable 
description of CARs 
has to go beyond formal-kinetic considerations and to 
take the influence of higher-order
 particle correlations into account. Here we develop such an approach.

We start from the following  system: The catalytic substrate consists of 
$N$,
immobile
catalytic sites (CSs), randomly placed 
in the
$d$-dimensional reaction volume  $V$. The reacting particles of type $A$ are 
also randomly positioned;  their initial 
average density  is 
$n_{0} = n_{A}(0)$. 
The $A$-particles  diffuse with  diffusion
coefficient  $D$.   
For simplicity we neglect sterical hindrances between the $A$s and also between
$A$s and CSs. 
Furthermore, 
we assume that 
the $A$ 
particles do not get trapped by the CS to form
metastable  $A$-CS complexes. 
Next, 
pairs of  $A$ particles may
react upon encounter, if they meet at a CS (Fig.1). Formally,  the elementary reaction step is
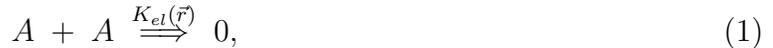
\begin{equation}
 A \; + \; A \; \stackrel{K_{el}(\vec{r})}{\Longrightarrow} \; 0,   
\end{equation}
where 
$K_{el}(\vec{r})$, is a random function of the 
spatial variable $\vec{r}$; 
$K_{el}(\vec{r})$ is strictly zero everywhere,  
except in the vicinity of any of the CSs,
in which case $K_{el}(\vec{r}) = K_{el}$. Here we take the CSs to be randomly distributed in $V$ 
with a constant 
average density $n_{C}$: thus the CSs' set and $V$ have the same topological dimension. We hasten to add,
however, that our approach  can be readily extended to other distributions of CSs (i.e. regular or
strongly inhomogeneous).

Since in $V$ many microscopic reactions which follow Eq.(1) go on at the same time, 
the global picture becomes complex, and is, in general, properly described
 by an infinite hierarchy of
coupled, differential equations$^{3,4}$. 
The analytical approach for CARs which we pursue here 
is 
 based on the truncation of this coupled system 
at the level of
 third-order correlations;  the latter 
  are then accounted for  through  appropriate
boundary conditions. For non-catalytic reactions in $3d$
this method
was pioneered by Collins and Kimball$^{8}$ (CK);  it generalizes 
Smoluchowski's treatment
of bimolecular chemical reactions 
(see, e.g. Ref.6).  The CK-approach provides 
 both for the
binary reaction $A + A \to 0$,  and for the
recombination reaction $A + B \to 0$ with $n_{A}(0) \neq n_{B}(0)$ 
  a reasonable description$^{8,9}$  of the experimentally observed kinetics
over the entire time domain. We note, however, that the reaction
$A + B \to 0$ with $n_{A}(0) = n_{B}(0)$ requires to go beyond the CK-approach, 
in order to depict the kinetic behavior at very long times, where  many-particle effects
come into play$^{4-6,10}$.

Two remarks are here appropriate:\\
(a) In the case when the CSs cover $V$ completely, 
 CARs behave exactly as non-catalytic reactions. This
special case  is 
 equivalent to the original CK-problem.\\
(b) Recently, 
the kinetics of the $A + B + C \to 0$ diffusion-limited reaction 
has been analysed$^{11,12}$ 
 using an extension
of the Smoluchowski approach (see, e.g. Ref.6). 
Setting $A = B = C$ leads formally to the reaction 
  $A + A +
A \to 0$, for which the procedure shows fair agreement$^{13}$  between the analytical
predictions and the numerical data.  For CARs 
 one may now be tempted to follow a similar course, by setting
 $A = B$ in $A + B + C \to 0$, and identifying  the Cs with the CSs.
We will show, however, that such a procedure does not describe CARs well; in
 $1d$ and in $2d$ 
 it does not lead to the proper 
long-time decay and in $3d$  it accounts 
only partially  for the effect of CSs.

The paper is structured as follows: In Section II we formulate the model,  by writing 
its 
basic equations, which allows us to  extend the $3d$ CK-approach to arbitrary $d$ and to
CARs. Here we also point out  the relation between the CK-CARs 
kinetics in $d$-dimensions and the trapping problem in $(d+d)$-dimensions.
In Section III we present explicit results for CARs'
 effective rates and for the reactants' decay; we compare these to the findings for 
non-catalytic
reactions.  Finally, we
conclude with a summary of results  in Section IV.

\section{The model and its basic equations.}

We formulate our model on a $d$-dimensional ($d = 1, 2, 3)$ 
 lattice with lattice spacing $a$.
 To each site of the lattice, whose position 
is specified by the
vector  $\vec{r}$,  we assign a time-independent
variable $n_{C}(\vec{r})$ which assumes
two possible values, namely  $0$ or $1$, depending on whether the site
is catalytic,
 $n_{C}(\vec{r}) = 1$, or not,
$n_{C}(\vec{r}) = 0$.  
The catalytic
substrate is the set of all $N$ CSs; 
we  denote 
it
by $\{\vec{R}_{k}\}$, where $\vec{R}_{k}$ is the vector of the $k$-th
CS. Here
we take the
$\vec{R}_{k}$ 
to be random, independent, uniformly distributed
 variables. The CSs density is  
 $n_{C} = V^{-1} \sum_{\vec{r}} n_{C}(\vec{r})$, where
the sum runs over
all lattice sites.  
The case when  $\{\vec{R}_{k}\}$ forms  different types of
ordered geometrical arrays
 will be discussed elsewhere.

We start at  $t = 0$  with randomly distributed, identical $A$
 particles, with mean density   $n_{0}$. Each $A$ particle moves by
  jumps to  nearest-neighboring  sites,   the average time interval 
 between 
 successive
jumps being  $\tau$. We disregard any excluded volume interactions; thus
 all $A$ particles perform independent random 
walks,  with the associated 
 diffusion coefficient $D = a^{2}/2 d \tau$.

Now,  whenever an $A$ particle lands on a 
 catalytic site which is already occupied by another $A$, 
the
 two $A$s may react 
at a   rate  $K_{el}$. Reacting $A$s are immediately removed 
from the system, whereas the corresponding
CS remains unaffected. On the other hand,  $A$s   never react
at  
non-catalytic sites.

\subsection{Evolution of the local density and of the two-point joint density
functions.}

Let $n(\vec{r},t)$ denote the local density of the $A$s. 
In continuous-time $n(\vec{r},t)$ obeys:
\begin{equation}
\left. \frac{d}{d t} \; n(\vec{r},t) \; = 
 \; \frac{1}{2 d \tau} \; \triangle_{\vec{r}} 
\; n(\vec{r},t) \; - 
\;  K_{el} \; n_{C}(\vec{r}) \; n(\vec{r}_{1}, \vec{r}_{2}; t)\right| _{\vec{r}_{1} = 
\vec{r}_{2} = \vec{r}}
\end{equation}
Here  $n(\vec{r}_{1}, \vec{r}_{2}; t)$ is the
two-point joint density function (i.e. 
 the probability of having at time
 $t$ an $A$
 at  $\vec{r}_{1}$ and another $A$ at  $\vec{r}_{2}$),  and 
the symbol $\triangle_{\vec{r}}$ 
stands for the following
 difference operator acting on  $\vec{r}$:
\begin{equation}
\triangle_{\vec{r}} \; n(\vec{r},t) \; =  \; - \; 2 \; d \; n(\vec{r},t) \; 
+ \; 
 \sum_{\vec{r} \; ^{\prime}, nn} 
n(\vec{r} \; ^{\prime},t)  
\end{equation}
Here the sum runs over nearest neighbors only, 
i.e. $|\vec{r} \; ^{\prime} - \vec{r}| = a$.

Apart from the factor $ n_{C}(\vec{r})$ stemming from the CS,  Eq.(2) is 
the conventional rate equation for  $A + A \rightarrow 0$ 
 (see e.g. Ref.8 for a discussion). On the rhs of Eq.(2) 
the first 
term accounts
for the  particles' migration, while the second one
describes the reaction: In  standard fashion, the reaction term
is taken to equal the product of the rate $K_{el}$, 
of the probability of having a pair of $A$ particles 
at the same site and 
at the same time, and of  $n_{C}(\vec{r})$;  
the latter factor is new here and is due to the fact that 
 $A$s can react only at catalytic sites. 
We note that  Eq.(2)
embodies  mean-field assumptions: In a more rigorous approach one
has to use the three-body 
 probability that two $A$s
encounter each other at
 a CS;  this probability is here
decoupled by having it  represented  as the product of $n_{C}(\vec{r})$ 
and $n(\vec{r}, \vec{r}; t)$.

Before we turn to the analysis of 
the time evolution of $n(\vec{r}_{1}, \vec{r}_{2}; t)$, 
it is instructive to consider
the result of the simple kinetic approach, which later will serve 
 as reference. Taking the volume average of both sides of Eq.(2)
and assuming that the non-linear reaction term on the rhs of Eq.(2)
decomposes  
into a product of 
averaged local densities, we arrive at  the standard, formal-kinetic
 "law of mass action" (see, e.g. Ref.7)
\begin{equation}
\frac{d}{d t} \; n_{A}(t) \; = \;
  - \; K_{el} \; n_{C} \; n^{2}_{A}(t) 
\end{equation}
Here $n_{A}(t) = V^{-1} \sum_{\vec{r}} n(\vec{r},t)$ 
is the average density of the $A$s. From
Eq.(4) the decay of $n_{A}(t)$ at sufficiently long times, 
$t \gg 1/n_{C} K_{el} n(0)$, 
 follows:
\begin{equation}
n_{A}(t) \; \approx \; (n_{C} K_{el} t)^{-1} 
\end{equation}
It is important to note that $n_{A}(t)$ given by Eq.(5) is
independent of the spatial dimension and  is  the same 
in, say, $1d$ and $3d$.
 Second, the formal-kinetic approach
predicts that the 
effective reaction rate constant is
$n_{C} K_{el}$, 
 and is thus independent of other parameters, such as, e.g. the
particles' diffusivities.

We now turn to the analysis of 
 the time evolution 
of $n(\vec{r}_{1}, \vec{r}_{2}; t)$. We find that this
 quantity obeys (see, e.g. Ref.13)
\begin{equation}
\frac{d }{d t} \; n(\vec{r}_{1}, \vec{r}_{2}; t) \; =
 \; \frac{1}{2 d \tau} \; 
\{\triangle_{\vec{r}_{1}} \; + \; \triangle_{\vec{r}_{2}}\} \; n(\vec{r}_{1}, 
\vec{r}_{2}; t) \; 
 + \; K_{el} \; T, 
\end{equation}
in which the  terms in  curly brackets stem 
 from the  particles' motion, 
while
$T$ is a combination of joint three-point  distributions and 
arises due to  the reaction between the $A$s. 

Eqs.(2) and (6) are the first two equations of 
an infinite hierarchy of coupled differential equations (CDE). Such a
hierarchy of CDE cannot be solved exactly; 
in order to compute the evolution of $n(\vec{r},t)$ and thus of its mean value 
$n_{A}(t)$ one has to resort to some approximate
 methods.

\subsection{The Collins-Kimball's $3d$ problem and its extension to arbitrary $d$.}

Here we continue  by recalling Collins and Kimball's (CK) 
 analysis$^{8}$ of  
reaction kinetics: As discussed, their problem is identical to $3d$ CARs  in a 
"completely" catalytic medium, i.e. such that $n_{C} = 1$. 
In Ref.8 
   the hierarchy of CDE is  truncated at the level of the third-order
 correlations, i.e. 
 $T$ in Eq.(6) is set  to zero,  and the reaction between particles is accounted for  
by introducing a
 "radiation" boundary condition on $n(\vec{r}_{1}, \vec{r}_{2}; t)$; CK stipulate
that the local reaction rate at any point $\vec{r}$, i.e.  $K_{el} \; n(\vec{r}, 
\vec{r}; t)$,
should be exactly equal to the diffusive current of pairs of $A$ particles into this
 point. 
Accompanied by the  following boundary and initial conditions:
\begin{equation}
\left. n(\vec{r}_{1}, \vec{r}_{2}; t)\right| _{|\vec{r}_{1} - \vec{r}_{2}| \to \infty } 
\; \rightarrow \;
n^{2}_{A}(t),   
\end{equation}
and
\begin{equation}
n(\vec{r}_{1}, \vec{r}_{2}; t = 0) \; = \; n^{2}_{A}(0) \; = \; n^{2}_{0},   
\end{equation}
which signify that correlations in the particles' positions vanish at large
separations and that initially the particles are uniformly distributed in $V$, 
the CK-approach results in a closed system of linear equations; these  allow then
to calculate the
effective reaction rate and  the time evolution of $n_{A}(t)$.
We note that in their original paper$^{8}$ CK have only
considered 
 the $3d$ case. 
Clearly, however, in the formulation of Eq.(6), extending the CK-approach 
 to systems of arbitrary
spatial dimension $d$ is straightforward, as we show in the following.

Solving Eq.(6) (with $T = 0$) subject to the radiation boundary condition
and the  conditions in Eqs.(7) and (8), 
CK find$^8$ that $n_{A}(t)$ obeys  in $3d$:
\begin{equation}
\frac{d }{d t} \; n_{A}(t) \; = \; - \; K_{eff}(t) \; n^{2}_{A}(t) 
\end{equation}
Eq.(9) is similar to the "law of mass 
action" displayed in Eq.(4), since for both the rhs are proportional to
the second power of $n_{A}(t)$, in agreement with the 
elementary reaction 
act. The difference between Eqs.(9) and (4) is that now 
 $K_{eff}(t)$ stands for
an effective (in general, time-dependent)
 rate coefficient,
 which arises mainly from the 
particles' diffusion.

Now  $K_{eff}(t)$ 
can be  expressed through
its Laplace-transform:
\begin{equation}
K_{eff}(\lambda) \; = \; \int^{\infty}_{0} dt \; \exp( - \lambda t) \; K_{eff}(t), 
\end{equation}
for which CK find, for identical reacting particles 
  in the limit $\lambda \ll D/r_{0}^{2}$,   where 
$r_{0}$ denotes the reaction radius:
\begin{equation}
\frac{1}{K_{eff}(\lambda)} \; = \; \frac{\lambda}{K_{el}} \; + 
\; \frac{\lambda}{8 \pi D r_{0}} 
\end{equation}
The second term on the rhs of
Eq.(11), i.e. $\lambda/8 \pi D r_{0}$, 
stems from  the Laplace-transformed 
Smoluchowski constant;  the latter equals  the  particles' current
towards the surface of a single, immobile, adsorbing $3$-dimensional
sphere of radius $r_{0}$. 

This result is actually
a special case of the general, $d$-dimensional expression 
for $K_{eff}(\lambda)$, which can be derived within the framework of 
 the CK approach, as we briefly outline now.

Consider the behavior of the two-point joint density function $n(\vec{r}_{1}, \vec{r}_{2}; t)$. 
Within the CK approach it
 obeys
 Eq.(6) with $T = 0$:
\begin{equation}
\frac{\partial }{\partial t} \; n(\vec{r}_{1}, \vec{r}_{2}; t) \; =
 \; D \; 
\{\triangle_{\vec{r}_{1}} \; + \; \triangle_{\vec{r}_{2}}\} \; n(\vec{r}_{1}, 
\vec{r}_{2}; t), 
\end{equation}
where $\triangle_{\vec{r}_{i}}$ denote the continuous-space Laplace operators. Note that Eq.(12)
is a  $(d+d)$-dimensional diffusion
equation. 
Now, 
Eq.(12)
is to be solved subject to
the initial and boundary conditions of Eqs.(7) and (8), as well as to the CK radiation
boundary condition
\begin{equation}
\left. K_{el} \; n(\vec{r}_{1}, \vec{r}_{2}; t) \; = \; D \; 
\{\nabla_{\vec{r}_{1}} \; + \; \nabla_{\vec{r}_{2}}\} \; n(\vec{r}_{1}, 
\vec{r}_{2}; t)\right| _{|\vec{r}_{1} - \vec{r}_{2}| 
= r_{0}} 
\end{equation}
As already observed by CK for non-catalytic reactions, $n(\vec{r}_{1}, \vec{r}_{2}; t)$
 depends only on $z = |\vec{r}_{1} - \vec{r}_{2}|$, 
the
relative distance between $\vec{r}_{1}$ and $\vec{r}_{2}$, 
and consequently, 
 the $(d+d)$-dimensional  Eqs.(12) and (13) reduce to the effectively $d$-dimensional equations:
\begin{equation}
\frac{\partial }{\partial t} \; n(z; t) \; =
 \; 2 \; D \; \{\frac{\partial^2}{\partial z^2} \; + \; 
\frac{d - 1}{z} \; \frac{\partial}{\partial z}\} \; n(z; t), 
\end{equation}
and
\begin{equation}
\left. K_{el} \; n(z; t) \; = \; 2 \; 
D \; \sigma_{d} \; \frac{\partial  n(z; t)}{\partial z}\right| _{z = r_{0}}
\end{equation}
Here $\sigma_{d} = 2 \pi^{d/2} r_{0}^{d - 1} \Gamma^{-1}(d/2)$ 
denotes the surface area of the $d$-dimensional sphere of radius $r_{0}$, with
 $\Gamma(x)$ being the Gamma-function \cite{abra}.
Equations (14) and (15)  generalize 
to arbitrary   dimensions the $3d$ equations studied by CK$^8$. 

Now, the 
solution of Eq.(14) subject to the 
 conditions of Eqs.(7), (8) and (15) can be
readily obtained by  Laplace transforming with respect to $t$.
 The Green's function $G_{d}(z;\lambda)$ of the equation
\begin{equation}
\lambda \; G_{d}(z; \lambda) \; =
 \; 2 \; D \; \{\frac{\partial^2}{\partial z^2} \; + \; 
\frac{d - 1}{z} \; \frac{\partial}{\partial z}\} \; G_{d}(z; \lambda) 
\end{equation}
obeys
\begin{equation}
G_{d}(z;\lambda) \; = \; \frac{2}{(8 \pi D)^{d/2}} \; (\frac{z^2}{8 D \lambda})^{(2 - d)/4} \;
\tilde{K}_{1 - d/2}(\sqrt{\frac{\lambda}{2 D}} z), 
\end{equation}
in which $\tilde{K}_{\nu}(x)$ is the modified Bessel function$^{14}$. 
The solution of the Laplace
transformed  Eq.(14), (note the occurence of an additional term, $- n_{0}^{2}$, on 
the lhs, as compared to Eq.(16)),
 has the form: 
\begin{equation}
n(z; \lambda) \; = \; \frac{n_{0}^{2}}{\lambda} 
\; [1 \; + \; A(\lambda) \; G_{d}(z;\lambda)], 
\end{equation}
in which the conditions Eqs.(7) and (8) have already been incorporated. In Eq.(18) 
 $A(\lambda)$ is a constant, which has to be chosen in such a way that the boundary
condition, Eq.(15), is also satisfied. 
Substituting Eq.(18) into Eq.(15) we find that $A(\lambda)$ is:
\begin{equation}
A(\lambda) \; = \; - \; G_{d}^{-1}(r_{0}; \lambda) \; 
\{1 \; - \; \frac{2 D \sigma_{d}}{K_{el}} \; [ln G_{d}(r_{0}; \lambda)]'\}^{-1}, 
\end{equation}
where the prime stands for the derivative with respect to $r_{0}$. 

This provides a complete solution of the 
CK-problem in $d$-dimensions.  In order to calculate the effective rate
constant, $K_{eff}(\lambda)$, we have merely to insert Eq.(18) with (19) into
 the rhs of Eq.(15). This yields
\begin{equation}
K_{eff}(\lambda) \; = \; - \;  \frac{2 D \sigma_{d}}{\lambda} \; [ln G_{d}(r_{0})]' \; 
\{1 \; - \; \frac{2 D \sigma_{d}}{K_{el}} \; [ln G_{d}(r_{0})]'\}^{-1} 
\end{equation}
Equation (20) takes a  physically more revealing form, if we rewrite it
 as
\begin{equation}
\frac{1}{K_{eff}(\lambda)}  \; = \; \frac{\lambda}{K_{el}} \; + 
\; \frac{1}{K_{d}(D; \lambda)}, 
\end{equation}
where 
\begin{equation}
K_{d}(D; \lambda) \; = \; - \; \frac{2 D \sigma_{d}}{\lambda} \; [ln
G_{d}(r_{0})]' \; = \; \frac{2 D \sigma_{d}}{\lambda} 
\; (\frac{\lambda}{2 D})^{1/2} \;
\frac{\tilde{K}_{d/2}(\sqrt{\frac{\lambda}{2 D}}
 r_{0})}{\tilde{K}_{1-d/2}(\sqrt{\frac{\lambda}{2 D}} r_{0})} 
\end{equation}
Eq.(21) resembles the electrostatic law of addition of parallel resistances  and
displays the combined effect of two controlling factors - the
effect of the elementary reaction act and the effect of the transport
of particles towards each other.
Notice now that for $K_{el} = \infty$ the rate 
 $K_{d}(D; \lambda)$
 is 
the Laplace transform of the diffusive current towards the surface of a
single immobile, perfectly adsorbing $d$-dimensional sphere of radius $r_{0}$. Consequently,
$K_{d}(D; \lambda)$ is the $d$-dimensional
analog of the Smoluchowski constant and  Eq.(21) represents the desired generalization of Eq.(11)
to the $d$-dimensional case. We hasten to remark that Eq.(21) has already been obtained in Ref.10 in a 
different framework, based on the analysis of the third-order joint density functions;  Eq.(21) 
 also follows
from the general results of Ref. 15, which considered reversible reactions.

The behavior of $K_{d}(D; \lambda)$  depends in a fundamental way 
on the spatial dimension $d$.
 As is well-known (see e.g. Ref.9), 
in low dimensions ($d \leq 2$) $K_{d}(D; \lambda)$ tends to zero 
for $\lambda \to 0$ ($t \to \infty$). 
To be explicit, in the limit $\lambda \ll D/r_{0}^{2}$
 the parameter $K_{d}(D; \lambda)$ determined by Eq.(22) 
equals $(8 D/\lambda)^{1/2}$ for $d = 1$ and
$8 \pi D /( \lambda \; ln(8 D/ r_{0}^{2} \lambda))$ for $d = 2$. In higher dimensions
($d > 2$) $K_{d}(D; \lambda)$ approaches constant values at long times; the values for 
$K_{d}(D; \lambda)$, for, say $d = 3, 4$
 and $6$ turn out to be $8 \pi D r_{0}/\lambda$, $8 \pi^{2} D r_{0}^{2}/\lambda$ and 
$8 \pi^{3} D r_{0}^{4}/\lambda$,  
respectively. 
The dependence of $K_{eff}(\lambda)$ on $d$,
arising due to the second term in Eq.(21), leads, especially for $d \leq 2$, to deviations of the
decay laws from Eq.(5).

\subsection{The CK-approach extended to CARs.}

After this overview of the CK approach we now extend it to CARs.
Following Ref.8  we truncate
the hierarchy at the level of the third-order joint density functions, 
which yields
the continuous-space Eq.(12). 
Next, in our system only the
encounters which happen on a CS may lead to a reaction. 
Thus, the radiation boundary condition
is to be imposed  on the CSs only. Denoting by 
 $S_{k}$ the surface of  the $k$-th CS,  we hence have that instead of Eq.(13) the following boundary
condition should be imposed on $n(\vec{r}_{1}, \vec{r}_{2}; t)$:
\begin{equation}
\left.  K_{el} 
\; n(\vec{r}_{1}, \vec{r}_{2}; t)\right| _{\vec{r}_{1}, \vec{r}_{2} \in
S_{k}} 
\; =   \; D \; \left. 
\{\nabla_{\vec{r}_{1}} \; + 
\; \nabla_{\vec{r}_{2}}\} \; n(\vec{r}_{1}, \vec{r}_{2}; t)\right| _{\vec{r}_{1}, \vec{r}_{2} 
\in S_{k}}
\end{equation}
For simplicity  we take in the continuum $S_{k}$ to be
the surface of the   $d$-dimensional sphere of radius $r_{0}$ centered at $R_{k}$. 
Equations (12) and (23) 
are complemented by
the initial condition, Eq.(8), and by the boundary condition, Eq.(7); this 
constitutes 
a closed system of
linear equations which allows the computation of $ n(\vec{r}_{1}, 
\vec{r}_{2}; t)$. In turn, knowing $ n(\vec{r}_{1}, 
\vec{r}_{2}; t)$ and 
using Eqs.(2) and (23), we  have 
\begin{equation} 
\frac{d }{d t} n_{A}(t) \; = 
 \; - \left. \; \frac{D}{V} \sum_{k} \{\nabla_{\vec{r}_{1}} \; + 
\; \nabla_{\vec{r}_{2}}\} \; n(\vec{r}_{1}, \vec{r}_{2}; t)\right| _{\vec{r}_{1}, \vec{r}_{2} 
\in S_{k}},
\end{equation}
which defines the evolution of  
the property of interest, namely of $n_{A}(t)$,  the mean density 
of $A$s.

\subsection{Relation between  CARs 
and the trapping reaction.}

We note now  that it is expedient to view Eqs.(12), (23), (7) and (8)
from a somewhat different perspective,  which 
will allow  us to find eventually 
an approximate analytical solution for them,  and to explain, 
on simple physical grounds, some seemingly surprising
results. As a matter of fact, what 
the CK approach enables us to do 
it is to reduce
the problem of computing  the  rates of  binary reactions, taking
place in  $d$-dimensional
catalytic systems, to the analysis of (imperfect) 
trapping in 
 $(d+d)$-dimensional systems. 
The latter problem, and especially 
its quantum mechanical counterpart,
the  scattering of  quantum
 particles by immobile impurities,
have been extensively investigated  (see, e.g. Refs.15-18
 and references therein).

The relation between the CAR and the trapping problem 
can be most simply
 illustrated for the  binary CAR in $d = 1$. 
Note
that   Eqs.(12) and (23) 
describe  the evolution of the local density
of some compound particles,
moving with
 diffusion 
coefficient$^{19}$
$D$ 
 on a two-dimensional plane
 $(r_{1}, r_{2})$, where $r_{1}$ and $r_{2}$ 
are scalar variables. The particles may disappear at the locations 
$(R_{k},R_{k})$  
of the traps,   
placed  on the 
diagonal  $r_{1} = r_{2}$ (Fig.2).
Physically, each compound particle is 
 a pair 
of $A$ particles, 
whose coordinates on the one-dimensional line 
are $r_{1}$ and $r_{2}$ respectively; consequently,  
$r_{1} = r_{2} = R_{k}$ are the only points 
where two $A$s may enter into reaction, 
in which case 
the compound particle may 
be  destroyed   by the trap at
$R_{k}$ with a finite probability related to
$K_{el}$. 
In this language, the reaction rate of the CAR, i.e. the rhs of 
Eq.(24), attains  a quite lucid meaning: It equals the
volume-averaged diffusive 
current of compound particles through the $S_{k}$. 
One can now note the fundamental  
 distinction between 
 systems with $n_{C} = 1$ and with $n_{C} < 1$:
The former case 
corresponds to a situation in which the
traps cover the diagonal  $r_{1} = r_{2}$ completely; here
$n(r_{1}, r_{2})$ depends  only on the relative distance
between $r_{1}$ and $r_{2}$, i.e. one has $n(r_{1}, r_{2})
 = n(r_{1} - r_{2})$, which then reduces the problem to $1d$. On the other
hand,  
in the case $n_{C} < 1$ the traps fill only some
portion of the diagonal, and thus 
the kinetics remains
essentially two-dimensional, since $n(r_{1}, r_{2})$ depends
 on both spatial variables;
compound
particles can cross  the diagonal 
harmlessly, i.e. with zero reaction probability,
through the gaps between the traps. 

Returning now to 
the general problem of CARs in 
$d$-dimensional media,  the corresponding mapping leads  to considering a trapping problem
 involving
compound particles diffusing in a $(d+d)$-dimensional space in the presence of 
imperfect traps placed on a $d$-dimensional substrate.

Several analytical approaches
have been developed to describe the
kinetics of trapping  in systems with  non-uniform
 spatial
 distributions of
traps (see, e.g. Refs.6 and 15).  
We will search for the solution of Eqs.(12) and (23) in the spirit of 
the Green's function method$^{17}$. Here we
  merely outline the
steps involved,  and address the reader for more details to
 Refs.15 and 17. 
 
One starts with the Laplace-transformed  Green's function solution 
$G_{d+d}(\vec{\rho};\lambda)$ of 
the $(d+d)$-dimensional
diffusion, Eq.(12), 
\begin{equation}
G_{d+d}(\vec{\rho};\lambda) \; = \; \frac{2}{(4 \pi D)^{d}} 
\; (\frac{|\vec{\rho}|^{2}}{4 D \lambda})^{(1 - d)/2}
\; \tilde{K}_{1 - d}(\sqrt{\frac{\lambda}{D}} | \vec{\rho} |),  
\end{equation}
in which 
$\vec{\rho} = (\vec{r}_{1},\vec{r}_{2})$. 
Furthermore,  the
Laplace-transform of $n(\vec{r}_{1}, \vec{r}_{2}; t)$ is represented 
as a series in which each term is 
the Green's function solution of Eq.(12), 
centered around the position
 of the $k$-th  trap,
\begin{equation} 
n(\vec{r}_{1}, \vec{r}_{2}; \lambda) \; = 
\; \frac{n^{2}_{0}}{\lambda} \; + \; 
\sum_{k} C_{k}(\lambda) \; G_{d+d}(\vec{\rho} -  \vec{\Theta}_{k};\lambda), 
\end{equation}
with  the $d+d$-dimensional vector $\vec{\Theta}_{k} =
(\vec{R}_{k},\vec{R}_{k})$. 
Eq.(26) obeys automatically Eqs.(12),(7) and (8); 
the coefficients $C_{k}(\lambda)$ are to be choosen in such a way that Eq.(23)
is satisfied. 
Substituting Eq.(26) into Eq.(23) we arrive at the following system of
$N$ linear equations for the $C_{k}(\lambda)$:
\begin{equation}
-  \frac{n^{2}_{0}}{\lambda} \; = \; \{K_{el}^{-1} 
+ G_{d+d}(r_{0}; \lambda)\} \; C_{j}(\lambda) \; + 
 \; \sum_{k} { }^{\prime} \; C_{k}(\lambda) \;
G_{d+d}(\vec{\Theta}_{k} - \vec{\Theta}_{j}; \lambda),  
\end{equation}
where $j = 1, ...  , N$, and 
the prime indicates that the sum in Eq.(27) runs over all 
$k$ with the exception of $k = j$.

The exact solution of Eqs.(27) for a given distribution 
of $\{\vec{R}_{k}\}$ requires the
inversion of the random matrix $||G_{d+d}(\vec{\Theta}_{k}
 - \vec{\Theta}_{j}; \lambda)||$, see Ref.17. 
Neglecting
fluctuations 
in the distribution of the CSs, in which case 
the Eqs.(27) simplify considerably,  one 
obtains
\begin{equation}
C(\lambda) \; \approx \; - \; \frac{n^{2}_{0}}{\lambda 
\; \{ K_{el}^{-1} + G_{d+d}(r_{0}; \lambda) + M_{scr}\}},  
\end{equation}
where $M_{scr}$ denotes the screening integral (or "shielding" integral
in the formulation of  Ref.15) 
\begin{equation}
M_{scr} \; = \; < \sum_{k} { }^{\prime} \; 
G_{d+d}(\vec{\Theta}_{k} - \vec{\Theta}_{j}; \lambda) > \; \approx 
 \; n_{C} \; \int \; \int \; d\vec{r}_{1} d\vec{r}_{2} 
\; \delta(\vec{r}_{1} - \vec{r}_{2}) \; 
G_{d+d}(\vec{\rho}; \lambda) 
\end{equation}
In Eq.(29)
 the brackets denote averaging over the distribution of
$\{\vec{R}_{k}\}$, and
the integrations with respect to the variables $\vec{r}_{1}$
 and $\vec{r}_{2}$ extend over the whole
 volume occupied by CSs, excluding the volume of a 
$d$-dimensional sphere of radius
$r_{0}$. In the following we turn to the limit $N, V \to \infty$, while 
keeping the ratio $N/V$ fixed,  $N/V = n_{C}$.
 We note that Eq.(29)
is only approximate, since
 excluded-volume aspects
 between the CSs are neglected;
this limits the applicability of the expression to $n_{C}$
 sufficiently small.
We note also 
that within our CK-type description the dependence of the
effective reaction rate on the geometry of the catalytic substrate and/or the 
distribution
of the CSs enters only through the screening integral $M_{scr}$. 
Consequently, any other 
geometry of the substrate (it can be, for instance, a two-dimensional 
convoluted surface of porous materials,  
imperfect crystallites with broken faces, kinks and steps, or
polymers in solution$^1$)  
can be accounted for by the use of the appropriate distribution
functions and by corresponding  integrations  in Eqs.(29). 
In particular, the details of the 
averaging
procedure
in  the case when the integrations extend over Gaussian 
polymer chains in solution have been discussed
in Ref.20.

\section{Results.}

Now, combining Eqs.(26),(28) and (29), we find from  Eq.(24)
 that $n_{A}(t)$ obeys the effective "law of mass action" in Eq.(9). In the limit
$\lambda \ll \lambda_{D} = D/r_{0}^{2}$, ($t  \gg 
\tau_{D} = r_{0}^{2}/D$), when $M_{scr}$ and $G_{d+d}(r_{0};\lambda)$
reduce to 
$M_{scr} \approx n_{C} \; (\lambda \; K_{d}(D;\lambda))^{-1}$
and $G_{d+d}(r_{0};\lambda) \approx (\lambda \; K_{d+d}(D/2;\lambda))^{-1}$,  
the effective rate constant
attains for $n_{C} r_{0}^d \ll 1$ the form
\begin{equation} 
\frac{1}{K_{eff}(\lambda)} \; = \; \frac{\lambda}{n_{C} K_{el}} \; + 
\; \frac{1}{K_{d}(D; \lambda)}
\; + \; \frac{1}{n_{C} K_{d+d}(D/2;\lambda)},   
\end{equation}
which represents the desired generalization of the CK-type result, 
Eq.(21), to  CARs. 
Equation (30) is the main result of our analysis and allows
 to compute  $n_{A}(t)$, 
which is related to $K_{eff}(t)$ through Eq.(9). Hence: 
\begin{equation}
n_{A}(t) \; = \; n_{0} \; \{1 \; + 
\; n_{0} \; \int^{t}_{0} d\tau \; K_{eff}(\tau)\}^{-1}
\end{equation}

Let us consider first the $3d$ case. We recall  the
explicit forms of the parameters $K_{d}(D; \lambda)$, 
presented in the text
after Eq.(22), so that in $3d$  Eq.(30) takes the 
following form ($n_{C} r_{0}^3 \ll 1$):
\begin{equation}
\frac{1}{K_{eff}(\lambda)} \; = \; \frac{\lambda}{n_{C} K_{el}} \; + 
\; \frac{\lambda}{8 \pi D r_{0}}
\; + \; \frac{\lambda}{4 \pi^{3} D r_{0}^{4} n_{C}},  
\end{equation}
which signifies that in the limit $t \to \infty$
 the effective rate constant
$K_{eff}(t)$ approaches a constant value:
\begin{equation}
\frac{1}{K_{eff}} \; = \; \frac{1}{n_{C} K_{el}} \; + 
\; \frac{1}{8 \pi D r_{0}}
\; + \; \frac{1}{4 \pi^{3} D r_{0}^{4} n_{C}}   
\end{equation}
Consequently, from Eq.(31) we have for $n_{A}(t)$ in three-dimensions
and large $t$:
\begin{equation}
n_{A}(t) \; \approx \; (K_{eff} \; t)^{-1},  
\end{equation}
where $K_{eff}$ is given by Eq.(33). 
Equation (34) signifies that in $3d$ catalytic
systems $n_{A}(t)$ is inversely proportional to $t$, i.e. 
is qualitatively the same as the formal-kinetic Eq.(5). Now, Eq.(33)
differs from Eq.(5), since here $K_{eff}$ replaces $n_{C} K_{el}$. Note from
Eq.(33) that $K_{eff}$ reduces to $n_{C} K_{el}$ only for  $D \to
\infty$. 
Thus 
for finite $D$, the rate $K_{eff}$ 
depends both
 on the reaction radius and on the mean density
of CSs.  
The  difference between
the CK-result for $n_{C} r_{0}^3 \sim 1$ 
and for  CARs with $n_{C} r_{0}^3 \ll 1$ 
is the last term in Eq.(30), which is due to the Green's function solution
of the diffusion equation in 
$6d$.  For small values of the parameter 
$n_{C} r_{0}^{3}$ and for $D r_{0}^{4} \ll K_{el}$,
the last term in Eq.(30) provides the dominant 
contribution to the  effective rate constant $K_{eff}$. 
In this case
Eq.(30) reduces to the result of Ref.12, obtained for the three-body problem.

Consider next the evolution of $n_{A}(t)$ 
in low dimensional systems, i.e. for $d = 1$ and $d = 2$.
We have from Eq.(30) that in $1d$  $K_{eff}(\lambda)$ is given by ($n_{C} r_{0}  \ll 1$):
\begin{equation}
\frac{1}{K_{eff}(\lambda)} \; = \; \frac{\lambda}{n_{C} K_{el}} \; + 
\; (\frac{\lambda}{8 D})^{1/2}
\; + \; \frac{\lambda \; ln(4 D/\lambda r_{0}^{2})}{4 \pi D n_{C}} 
\end{equation}
Now, Eq.(35) shows that in $1d$ catalytic systems the kinetics is
richer than
in the $3d$ case: comparing the different terms in
Eq.(35) one readily notices that 
 depending on  $\lambda$ each of these terms may dominate
 $K_{eff}(\lambda)$; hence 
 a succession of different kinetic
regimes may be observed  in the time domain. 
When $4 \pi D/K_{el}$ is sufficiently large and $n_{C} r_{0}$
is sufficiently small, so that $n_{C} r_{0} \ll exp(-4 \pi D/K_{el})$, we can consider
three different intervals, namely
\begin{equation}
\lambda_{D} \; exp(-4 \pi D/K_{el}) \; \ll \; \lambda \; \ll \; \lambda_{D},
\end{equation}
\begin{equation}
\lambda_{D} \; (n_{C} r_{0})^{2} \; \ll \; \lambda \; \ll \; \lambda_{D}  \; exp(-4 \pi D/K_{el}), 
\end{equation}
and
\begin{equation}
\lambda \; \ll \; \lambda_{D} \; (n_{C} r_{0})^{2} 
\end{equation}
(a) In the regime described by Eq.(36) the main contribution to $K_{eff}(\lambda)$ comes
from the first term on the rhs of Eq.(35); in this kinetically-controlled regime 
$K_{eff}(t) = n_{C} K_{el}$ and hence coincides 
with the result of the formal-kinetic approach. This behavior persists until
$t_{kc} \approx \tau_{D} \;  exp(4 \pi D/K_{el})$, which can be rather large for 
 $D \gg K_{el}$.  On the other hand, 
such a behavior is unobservable for
$K_{el} \gg 4 \pi D$, exemplified by instantaneous
reactions in Eq.(1). 

(b) In the regime described by Eq.(37) 
 $K_{eff}(\lambda)$ is dominated by the
third term on the rhs of Eq.(35), so that $K_{eff}(t) \approx
4 \pi D n_{C}/ln(4 t/\tau_{D})$. This $1d$ CAR expression is reminiscent 
(apart of the factor
$n_{C}$ and the replacement $D \to D/2$) of 
the classical result of Ref.9 for
the kinetics of binary reactions in 
$2d$. One thus
expects that at this stage $n_{A}(t) \sim ln(t)/t$.
The appearence of such an effectively $2d$
regime for CAR in $1d$  constitutes 
the principal difference between the CAR kinetics for $n_{C} r_{0} \ll 1$ and the kinetics 
of noncatalytic binary reactions
in $1d$.  We also note 
that such an effectively $2d$ behavior
was predicted
in Ref.12 for $1d$ reactions of $A + B + C \to 0$ type
as the final kinetic stage; in our case,  when $C$ does not disappear
in the reaction act, Eq.(37) determines only an intermediate transient
stage, which may be observed for times smaller   than  a typical time $t_{tr}$, 
where  $t_{tr} = 1/D
n_{C}^{2}$. Consequently, for $1d$
systems with very low densities of catalytic
sites $(n_{C} r_{0} \ll 1$) such an effectively $2d$ behavior can last  over 
extended time periods.

(c) Finally, in the limit of very small $\lambda$, Eq.(38), $K_{eff}(\lambda)$ 
 is determined by the second term on the rhs of Eq.(35), i.e. $K_{eff}(\lambda) \approx (8
D/\lambda)^{1/2}$. 
Thus for large $t$ the rate $K_{eff}(t)  \approx (8
D/ \pi t)^{1/2}$ is 
independent of $n_{C}$. 
Hence for large times the qualitative decay behavior is the same for CARs 
and for non-catalytic reactions.
 Actually, this seemingly surprising behavior has already been observed 
 numerically$^{21}$ and has a simple physical interpretation, which we will discuss below.
Explicitly, we find that the mean density of $A$ particles decays in the limit $t \gg t_{tr}$ as
\begin{equation}
n_{A}(t) \; \sim \; (\pi/32 D t)^{1/2} 
\end{equation}

We turn next to the analysis of
 CAR in $2d$. Now Eq.(30) reads:
\begin{equation}
\frac{1}{K_{eff}(\lambda)} \; = \; \frac{\lambda}{n_{C} K_{el}} \; + 
\; \frac{\lambda \; ln(8 D/r_{0}^{2} \lambda)}{8 \pi D} \; + 
\; \frac{\lambda}{4 \pi^2 D r_{0}^{2} n_{C}} 
\end{equation}
On comparing different terms on the rhs of Eq.(40)
 we  infer that in $2d$ two different kinetic stages may take place. Namely, for $\lambda$
from the interval
\begin{equation}
\lambda_{D} \; exp( - \frac{8 \pi D}{n_{C} K_{el}} - \frac{1}{\pi n_{C} r_{0}^{2}})
 \; << \; \lambda \; << \; \lambda_{D},
\end{equation}
the sum of the first and of the third term, which both have the same $\lambda$-dependence,
determines $K_{eff}(\lambda)$, while for smaller $\lambda$, such that
\begin{equation}
\lambda \; << \; \lambda_{D} \; exp( - \frac{8 \pi D}{n_{C} K_{el}} - \frac{1}{\pi n_{C}
r_{0}^{2}}),
\end{equation}
the main contribution to the effective rate constant is given by the second term on the rhs of
Eq.(40). 

(a) In the regime described by Eq.(41) $K_{eff}(\lambda)$ is controlled by the
constraints imposed by the elementary reaction act and by the diffusion in 
$4d$. For such a regime we  obtain
\begin{equation}
n_{A}(t) \; \approx \;  (\frac{1}{K_{el}} \; + 
\; \frac{1}{4 \pi^2 D r_{0}^{2}}) \; (n_{C} t)^{-1} 
\end{equation}
According to Eq.(41), the  regime described by
Eq.(43) is a transient one and
persists until $t_{tr} \approx \tau_{D} \; exp((D/n_{C} K_{el}) +
(1/r_{0}^{2} n_{C}))$.

(b) In the final stage described by Eq.(42), the 
main contribution to $K_{eff}(\lambda)$
comes from the second term in Eq.(40). Thus,  
similarly to the behavior in $1d$, 
$K_{eff}(\lambda)$ in the limit $\lambda \to 0$ 
is independent of $n_{C}$. 
This implies that also in $2d$ for large times the
 kinetics for CARs is the same as for non-catalytic reactions.
 We find 
here 
\begin{equation}
n_{A}(t) \; \sim \; \frac{ln(8 D t/r_{0}^{2})}{8 \pi D t}, 
\end{equation}
which coincides with the result of Ref.9 obtained
 for the long-time kinetics of non-catalytic reactions.  
We furthermore note  that the very long time decay
 behavior  is
reached  much more slowly in $2d$ than in $1d$;  the crossover time
$t_{tr}$ is in $2d$ an exponential function of $n_{C}^{-1}$ and is
 thus substantially larger
than its  $1d$ counterpart, which goes as $n_{C}^{-2}$.

Lastly, we discuss the physical origin of the fact
 that in low dimensions the CARs long-time
decay is independent of $n_{C}$. Here,  
the analogy between
the CAR kinetics in $d$-dimensions and the trapping
problem in $(d+d)$-dimensions again turns out to be 
very fruitful.  Let us consider first
the case  $d = 1$.  The equivalent problem is a $d = 2$ system
with compound particles diffusing 
in the presence of traps placed on the
 diagonal $r_{1} = r_{2}$, (see Fig.2). 
Now, 
it is well-known$^{9}$ that even in the
 presence of a single trap the particles' 
density profile around the trap
is not stationary;  there is a zone around the trap, 
which is depleted of  particles
and whose size grows with time as  $\sqrt{t}$.  
For the situation depicted
in Fig.2 at short times (such that $n_{C} \sqrt{D t} \ll 1$)
 the depletion zones of different traps 
 are well separated
from each other: Consequently, at
 short times the traps act independently
and the Laplace-transformed particle current 
 towards a given trap is given by Eq.(21),  $J(\lambda) \approx
K_{el} K_{2}(D/2; \lambda)/(K_{el}  + K_{2}(D/2; \lambda))$. 
The effective rate,  which is given by Eqs.(23) and (24), is now 
$K_{eff}(\lambda)  = n_{C} J(\lambda)$; the CAR decay 
 shows  in  this time-domain
an effectively two-dimensional behavior going as 
$n_{A}(t)  \sim n_{C} ln(t)/t$. This behavior also
 shows up in  the results of Ref.12.
At longer times, the depletion zones of different traps start 
to overlap and the probability $n(r_{1},r_{2})$ of 
finding a compound particle
on the diagonal $r_{1} = r_{2}$ decreases
 substantially  even in the gaps between the traps.
This results
in a situation in which the array of traps distributed on the
 diagonal $r_{1} = r_{2}$ acts
as an adsorbing line. It is not surprising then that the 
current of particles per trap attains
a one-dimensional form $J(\lambda) \approx K_{d=1}(D; \lambda)/n_{C}$,  
and that 
even for $n_{C} r_{0} \ll 1$ at very long times the
evolution of $n_{A}(t)$ proceeds essentially in the same fashion 
as for non-catalytic reactions. Similarly,
 for $2d$ CARs we have to 
analyse the kinetics
of  trapping in $4d$,  the traps being now
distributed on a $2d$
plane. As before,  one can now  distinguish
 between two different temporal regimes: 
one finds first a situation  in which the
traps act independently, which gives rise to an effectively 
$4d$ behavior$^{12}$. This regime crosses over to a stage at long times, where 
 the depletion zones of the different traps overlap, so
 that the array of traps
acts as an effectively $2d$ adsorbing plane;  the decay of 
  $n_{A}(t)$ is then given by   Eq.(41), and is independent of $n_{C}$.
We note finally, 
that such a peculiarity of 
the trapping kinetics in
low dimensional system,  
associated with the formation
 of non-stationary depletion zones around
traps, has  already been demonstrated 
in Ref.15. In particular, it was
shown$^{15}$ that for $2d$ systems in
 which $N$ traps are located
inside a circular area of radius $R$ one finds two different temporal
regimes: In the first, intermediate time regime all traps act
independently and the effective rate constant is proportional to $N$. 
This regime crosses
over  into a long-time stage, in which
 the depletion zones around
different traps overlap; at this stage 
an array of traps acts as a single trap of
radius $R$ and the effective rate constant is independent of $N$.

\section{Conclusions.}

We now conclude with the discussion of the obtained results.
We find for $3d$ CARs that $n_{A}(t)$ 
decreases inversely proportional to $t$, Eq.(34),
which agrees with the formal-kinetic picture, Eq.(5). 
Distinct from it, the effective 
rate constant is
less than the formal-kinetic value, $n_{C} K_{el}$, and 
depends both on the particles' diffusion constant  $D$ and on the
reaction radius $r_{0}$. 
We also note  that in the diluted case, when 
$n_{C} r_{0}^{3} \ll 1$, the result
in Eq.(33) shows the same dependence on the system's parameters
 as the one predicted in
Ref.12.  The very  long-time
behavior in  low-dimensions  is somewhat
 surprising, since the decay turns out to be
essentially independent of  $n_{C}$.
The  approach to this asymptotic domain is,
 however, very slow for low densities
of CSs, $n_{C}  r_{0}^d \ll 1$, 
and thus different decay forms appear  
 at intermediate times. The crossover times
$t_{tr}$ may be very large, since we 
find that 
$t_{tr} \sim 1/n_{C}^{2}$ in $1d$ and $ln(t_{tr}) \sim 1/n_{C}$ in  $2d$. 
For $n_{C} r_{0}^d \ll 1$ and for extended period of
 time the decay laws obey  
in $1d$ and $2d$  
 $n_{A}(t) \sim n_{C} ln(t)/t$ and
$n_{A}(t) \sim n_{C}/t$  respectively.

\vspace{0.5cm} 

\begin{center}
\begin{Large}
Acknowledgments
\end{Large}
\end{center} 

The authors thank S.F.Burlatsky, J.Klafter, 
J.L.Lebowitz  and S.K.Nechaev
 for helpful discussions. We acknowledge the
support of the FNRS (Belgium), the CNRS (France),
 the DFG and 
 the Fonds der
Chemischen Industrie, as well as grants through
German-Israeli-Foundation-(GIF) and  PROCOPE-projects.

\pagebreak

\begin{Large}
Figure Captions
\end{Large}

\vspace{0.5cm}

Fig.1. Reactions in catalytic media. Open
 circles denote immobile catalytic sites;
 the filled circles stand for diffusive
$A$ particles. Case (1) shows a situation in
 which an encounter of $A$ particles
 does not lead to reaction,
while in the case (2) the reaction  may take place.

\vspace{0.5cm} 

Fig.2.  Open circles on the $r_{1}$ and $r_{2}$-axis 
denote the catalytic sites; 
filled circles give the
corresponding positions of traps and small black circles
 denote diffusive
 compound particles.


\begin{references}
\bibitem{ref} To generalize our mean-field-type approach to
the  analysis of the recombination reaction
$A + B \to 0$, we have merely to assume
anisotropic diffusion, where the diffusion coefficient 
of the compound particle
along the $r_{1}$-axis is $D_{A}$, while it
is $D_{B}$ along the $r_{2}$-axis.
\end{references}
\end{document}